\documentclass[preprint2]{aastex}

\usepackage{graphicx,epstopdf}

\usepackage{color}

\def\BB{{\bf {B}}}
\def\JJ{{\bf {J}}}

\def\vv{{\bf {v}}}

\def\XX{{\bf {X}}}

\def\ee{{\bf {e}}}
\def\nn{{\bf {n}}}

\shorttitle{Current layers in braided coronal loops}
\shortauthors{Pontin \& Hornig}

\begin{document}

%\title{On the degree of field line braiding in coronal loops}
%\title{The structure of current layers in braided coronal loops, and implications for coronal heating}
\title{The structure of current layers and degree of field line braiding in coronal loops}


\author{D.~I.~Pontin and G.~Hornig}
\affil{Division of Mathematics, University of Dundee, Dundee, UK}
\email{dpontin@maths.dundee.ac.uk}

\begin{abstract}
One proposed resolution to the long-standing problem of solar coronal heating involves the buildup of magnetic energy in the corona due to turbulent motions at the photosphere that braid the coronal field, and the subsequent release of this energy via magnetic reconnection. In this paper the ideal relaxation of braided magnetic fields modelling solar coronal loops is followed. A sequence of loops with increasing braid complexity is considered, with the aim of understanding how this complexity influences the development of small scales in the magnetic field, and thus the energy available for heating. It is demonstrated that the ideally accessible force-free equilibrium for these braided fields contains current layers of finite thickness.  It is further shown that for any such braided field, if a force-free equilibrium exists then it should contain current layers whose thickness is determined by length scales in the field line mapping. The thickness and intensity of the current layers  follow scaling laws, and this  allows us to extrapolate beyond the numerically accessible parameter regime  and to place an upper bound on the braid complexity possible at coronal plasma parameters. At this threshold  level the braided loop contains $10^{26}$--$10^{28}{\rm ergs}$ of available free magnetic energy, more than sufficient for a large nanoflare.

\end{abstract}



\section{Introduction}
The heating of the solar coronal plasma to multi-million degree temperatures remains one of the outstanding questions in solar physics, more than 70 years after the presence of this hot plasma was first realised. One of the leading theories proposed is Parker's \citep{parker1983,parker1988} nanoflare model, in which convective motions at the photosphere tangle and stress magnetic field lines in the corona, injecting energy that is eventually liberated in impulsive heating events facilitated by magnetic reconnection. In order for reconnection (a nanoflare) to be initiated, current sheets must form on extremely small length scales to allow reconnection to proceed in the highly conducting coronal plasma. Parker's original proposition was that for a sufficiently tangled field no corresponding smooth equilibrium exists, with the magnetic field relaxation instead leading to the formation of tangential discontinuities -- singular current sheets -- in the perfectly conducting limit \citep{parker1972}. This idea has attracted significant debate, with a number of studies arguing both for \citep[e.g.][]{ng1998,low2006,janse2010} and against \citep{vanballegooijen1985,longcope1994,craigsneyd2005} spontaneous current sheet formation. A review of field line braiding models is presented in \cite{wilmotsmith2014}. Of particular relevance to the present study are the results of \cite{vanballegooijen1988a,vanballegooijen1988b} and \cite{mikic1989}. In each paper a sequence of shear boundary displacements was applied to an initially homogeneous field between two line-tied plates (each representing a section of the photosphere), with the field allowed to relax to an equilibrium after each subsequent displacement. In both cases an absence of singular current structures was reported, although the authors observed an exponential increase of the peak current and exponential decrease in the current layer length scales with each successive boundary displacement. This provides an alternative scenario for nanoflare triggering, namely that continued tangling of the magnetic field by boundary motions leads to current layers of finite thickness on progressively smaller scales, with reconnection onset being triggered when the current density or current layer thickness reaches some threshold set by the local plasma conditions.

Whether the nanoflare model is capable of explaining the observed properties of loops in the active and quiet corona depends on a number of further outstanding issues. The coronal magnetic field must be able to store sufficient energy to provide the source of heating, and then the nanoflare mechanism must be capable of releasing a sufficient fraction of this energy to heat the plasma, on an appropriate timescale. One further crucial aspect to determining the efficacy of the nanoflare mechanism -- outwith the realm of the present study -- is to understand the plasma response to the energy deposition, see e.g.~\cite{cargill1997,cargill2014}. Indeed, one also needs to understand the temporal distribution of nanoflares, as well as their energy spectrum.
Extensive reviews of the observed properties of coronal loops \citep{reale2010} as well as broad aspects of the coronal heating problem \citep{klimchuk2006} are available.

We focus on understanding the role of field line braiding as a potential  trigger mechanism for a nanoflare. Numerous studies exist in which resistive MHD simulations of the braiding mechanism are performed directly, with time-dependent driving applied at simulation boundaries, and a subsequent heating of the plasma in the domain demonstrated \citep[e.g.][]{galsgaard1996,gudiksen2002,gudiksen2005,bingert2011}. However, such approaches must always employ a parameter regime that is some orders of magnitude off that of the corona, for numerical tractability. Most significantly, the Lundquist number is typically 10 orders of magnitude lower in the simulations than the actual value in the corona. Here we use an approach that excludes much of the complexity of the coronal dynamics, in which we do not treat the boundary driving explicitly. Rather, we assume that the magnetic field has already been braided by boundary motions, and investigate the relaxation of this  magnetic field towards equilibrium in the perfectly conducting limit. This simpler problem setup avoids the issue of magnetic reconnection setting in too early in the braiding process due to unrealistically  low magnetic Reynolds numbers.
%allows a detailed understanding to be gained into one aspect of the problem, and should be considered as being complimentary to the above studies. 

Specifically, we investigate the ideal MHD relaxation of a sequence of magnetic fields with increasing braid complexity, with the aim of understanding how this complexity influences the development of small scales in the magnetic field. 
%We demonstrate that this ideal relaxation does not lead to the formation of tangential discontinuities. Rather it leads to thin but finite current layers, the intensity of which increase, and the thickness decrease, with increasing magnetic field line complexity -- in agreement with \cite{vanballegooijen1988a,vanballegooijen1988b} and \cite{mikic1989}. We relate this directly to the complexity present in the magnetic field line mapping. 
%Extrapolating from these results allows us to predict the degree of braiding required to initiate a braiding-induced nanoflare at true coronal parameters. We then predict the expected properties of the magnetic field just prior to the nanoflare, such as the free energy.
In Section \ref{numsec} we describe the model setup and numerical methods. In Section \ref{ressec} we describe the results, and in Section \ref{discusssec} discuss their implications. We finish in Section \ref{concsec} with our conclusions.

\section{Model setup and numerical methods}\label{numsec}
\subsection{Model magnetic field}
%%%%%%%%%%%%
\begin{figure}
\centering
\includegraphics[width=3.5cm]{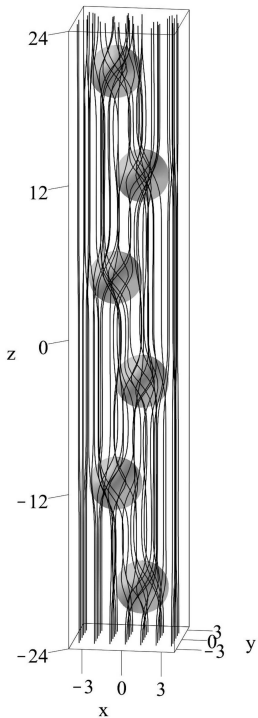}
\caption{Sample magnetic field lines from the initial magnetic field given by Eq.~(\ref{beq}) with $k=1$. Grey surfaces are isosurfaces of $|{\bf J}|$.}
\label{blines_ini}
\end{figure}
\begin{figure}
\centering
(a)\includegraphics[width=6cm]{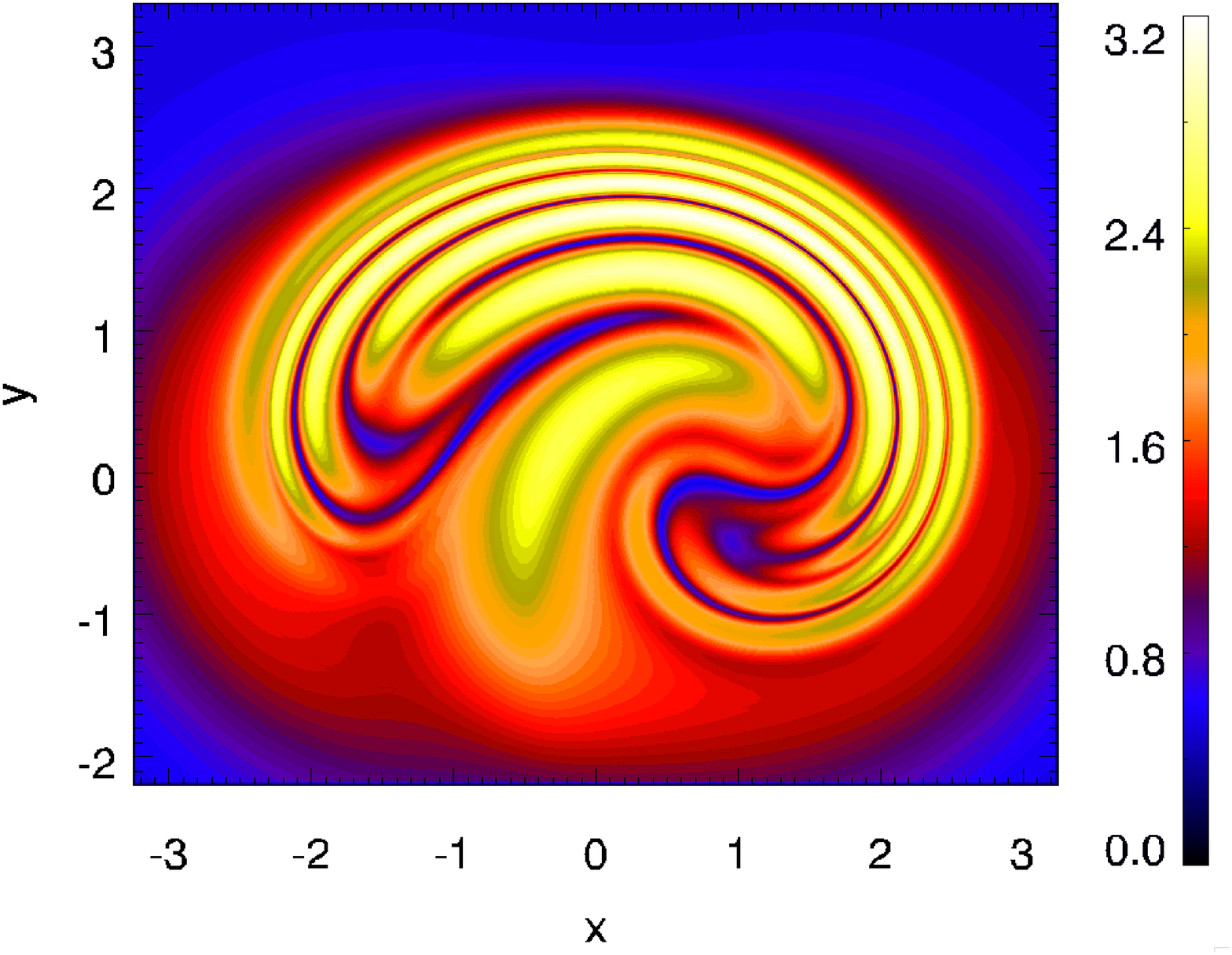}
(b)\includegraphics[width=6cm]{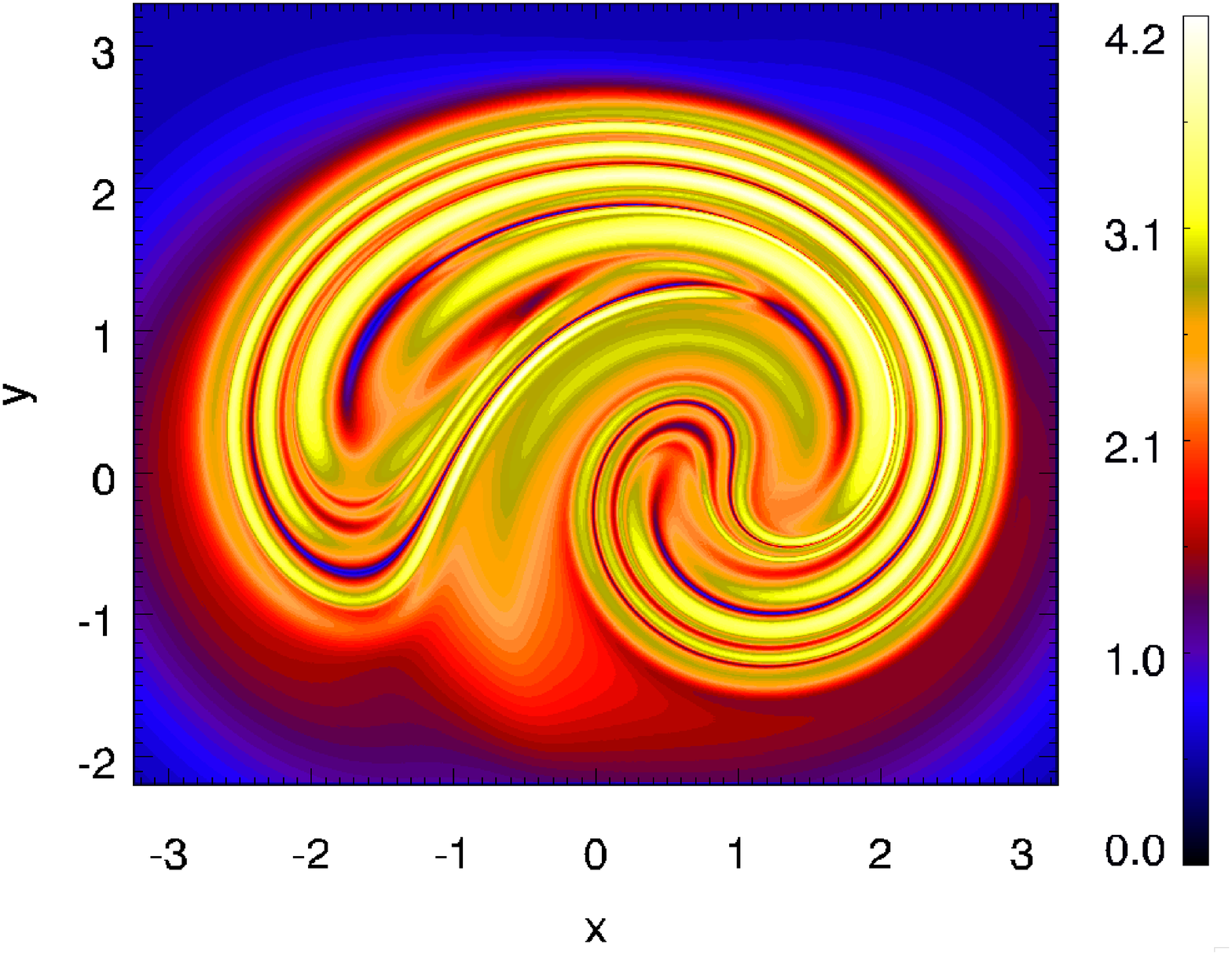}
(c)\includegraphics[width=6cm]{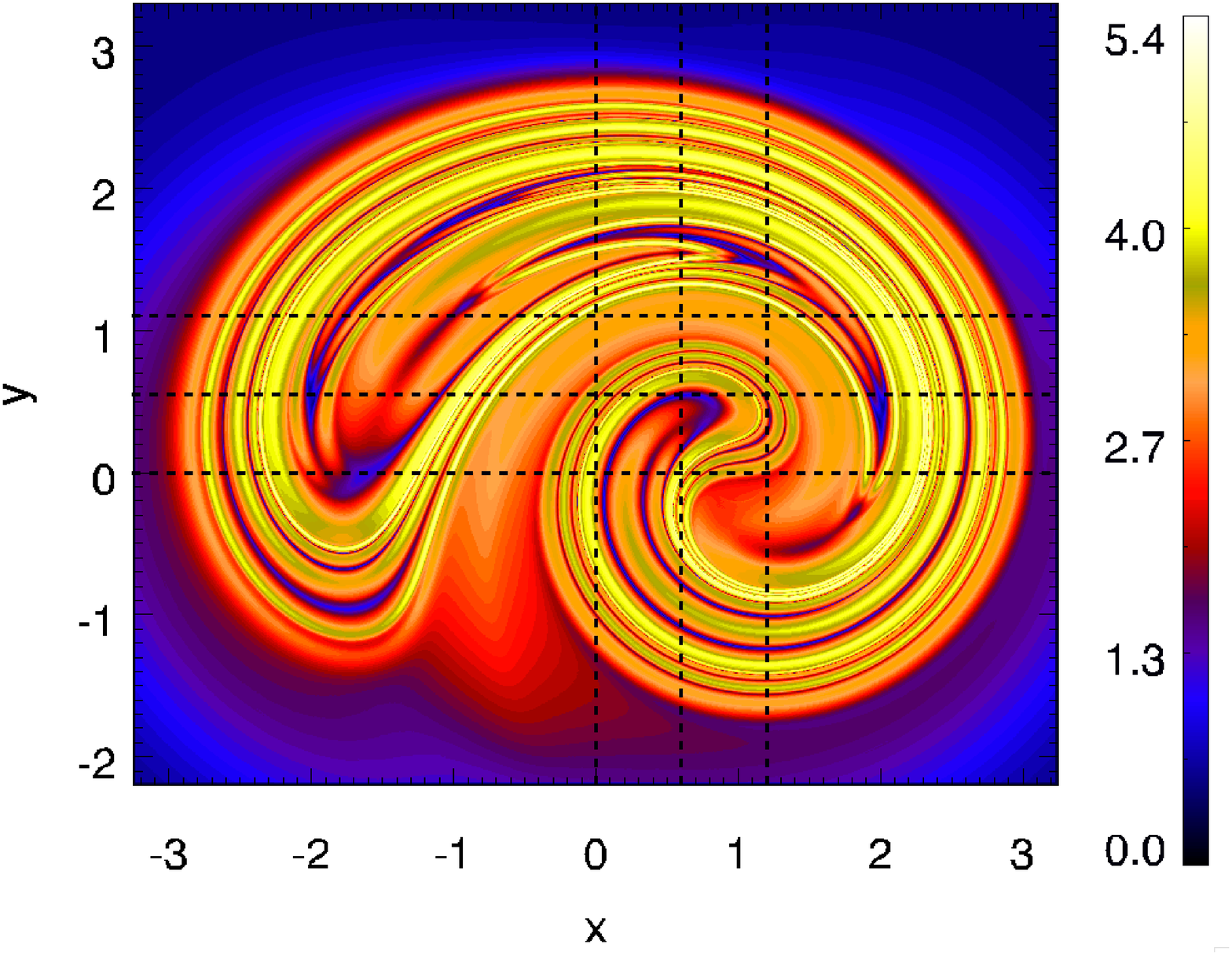}
\caption{$\log_{10}(Q)$, for squashing factor $Q$ defined in Eq.~(\ref{qeq}), for (a) $k=0.5$, (b) $k=0.75$, (c) $k=1$.}
\label{qmaps}
\end{figure}

%%%%%%%%%%%%
As a model magnetic field for our coronal loop we take
\begin{eqnarray}
\BB=\sum_{i=1}^{2n} k (-1)^i \exp{\left( -\frac{(x-x_i)^2+y^2}{2}-\frac{(z-z_i)^2}{4}\right)}\nonumber\\
\times (-y\,\ee_x+(x-x_i)\,\ee_y) ~+~B_0\ee_z \label{beq}
\end{eqnarray}
where we set $B_0=1$ and $n=3$ throughout, $x_i=(-1)^{i+1}$, and $z_{1..6}=\{-20,-12,-4,4,12,20\}$. Setting the parameter $k=1$, a subset of magnetic field lines in the domain have a `pigtail braid' structure, see \cite{wilmotsmith2009a} for further details. Magnetic field lines for $k=1$ are shown in Fig.~\ref{blines_ini}. More generally, for any value of $k$ field lines wind around one another in a non-trivial manner within the domain. This tangling of magnetic field lines along the loop generates a complex field line mapping between $z=-24$ and $z=24$, which represent here fragments of the photosphere. This mapping can be visualised by plotting the {\it squashing factor}, $Q$, defined by
\begin{equation}\label{qeq}
Q=\frac{\left(\frac{\partial X}{\partial x}\right)^2+\left(\frac{\partial X}{\partial y}\right)^2+\left(\frac{\partial Y}{\partial x}\right)^2+\left(\frac{\partial Y}{\partial y}\right)^2}{\left|\frac{\partial X}{\partial x}\frac{\partial Y}{\partial y}-\frac{\partial X}{\partial y}\frac{\partial Y}{\partial x}\right|},
\end{equation}
where $x$ and $y$ are field line footpoints on the `launch' boundary $z=-24$, and $X$ and $Y$ are the footpoint locations on the `target' boundary $z=24$, see \cite{titov2002,titov2007}. {The distribution of $Q$ is obtained by integrating field lines from a rectangular grid of typically around $10^7$ footpoints on $z=-24$ and then calculating the required derivatives using high-order finite differences (sixth-order centred differences) over this grid.} The squashing factor $Q$ is plotted as a function of the footpoint location on $z=-24$ for three different values of $k$ in Fig.~\ref{qmaps}. It is clear that increasing $k$ corresponds to increasing the topological complexity of the field, with $Q$ increasing in peak value and forming progressively thinner layers (usually termed {\it quasi-separatrix layers} or QSLs). We will return to discuss the thickness of these layers later. When we come to consider dimensional numbers, we will associate 1 length unit with $10^6m$. This gives a loop length of $48Mm$ and diameter (considering the region of braided flux) of around $6Mm$ (see Fig.~\ref{qmaps}) matching a typical, moderate-sized coronal loop.

Our purpose here is to investigate the implications of the complex field line mapping for the structure of the corresponding ideally-accessible force-free equilibrium. To this end, we perform a series of ideal relaxation simulations in which we take the magnetic field (\ref{beq}) as the initial condition, for different values of the parameter $k$.
%%%%%%%%%%%%
%%%%%%%%%%%%

\subsection{Numerical methods}
The ideal relaxation of the magnetic field described above is performed in two discrete stages. 
%The approach is similar to that described in {\bf a series of recent studies}, and we therefore provide here only a brief summary. 
Stage one  is performed using the ideal Lagrangian magnetofrictional relaxation scheme of \cite{craig1986}{ -- for a description of the implementation, see \cite{wilmotsmith2009a}}. {This numerical scheme has the desirable property} that it exactly preserves the magnetic topology (connectivity of all field lines in the domain). {However}, the disadvantage is that mesh distortions inhibit the accuracy of the method, and therefore restrict the proximity to force-free equilibrium that can be achieved \citep{pontin2009}. 
In order to further relax towards a force-free equilibrium the final state of this magnetofrictional relaxation is then taken as an initial condition for an ideal MHD simulation. This requires interpolating the vector potential from the deformed Lagrangian mesh onto a new rectangular mesh, which is performed using the method described in \cite{wilmotsmith2010}. The benefit of performing the magnetofrictional relaxation as a first step is that the majority of the free energy can be extracted at this stage (see below). This limits the magnitude of the plasma flows and dynamic magnetic fluctuations present in the MHD evolution, helping to minimise numerical dissipation. During both simulation stages the magnetic field is line-tied, i.e.~$\vv={\bf 0}$ and $\BB\cdot\nn$ fixed, on all boundaries.

The MHD code employed for the second stage of the relaxation is the Copenhagen Stagger Code\footnote{http://www.astro.ku.dk/$\sim$kg} \citep{nordlund1997,wilmotsmith2010}. A uniform viscous damping is employed to relax the field towards equilibrium, the corresponding term in the momentum equation being $\nu(\nabla^2\vv+\nabla(\nabla\cdot\vv)/3)$, where we take $\nu=0.01$ throughout. In contrast to the study of \cite{wilmotsmith2010}, the resistivity is set explicitly to zero, with no hyper-resistive terms included. Therefore the only changes of topology result from numerical dissipation, which is small due to the sixth-order derivative operators and high resolution. We demonstrate in Appendix \ref{appA} that the topological change due to numerical dissipation is negligible. 
%This is aided by the fact that the majority of the free energy in the ``pre-initial" field Eq.~(\ref{beq}) is extracted in the magnetofrictional relaxation. 
The plasma is initialised with a (non-dimensional) density of $\rho=1$ and thermal energy $e=0.01$, meaning that the plasma-$\beta\approx 0.013$. These MHD simulations are run for a domain of size $[12,8,48]$, centred at the origin, at a series of numerical resolutions; $180^2\times240$, $240^3$, $360^2\times 240$, $480^2\times 240$. {The simulations at different resolutions are used to check that fully resolved current layers are obtained (by checking for convergence of the current layer thickness with increasing grid resolution), described further below.}
%\tr{In addition, for $k=0.75$ one simulation run at $600^2\times 240$ is performed. } 

%%%%%%%%%%%%%%%%
%%%%%%%%%%%%%%%%
\section{Simulation results}\label{ressec}
In this section we describe the results of our simulations. For all values of $k$ the end state of the magnetofrictional stage  contains only large scale current structures similar to those discussed by \cite{wilmotsmith2009a}.  Hence we focus on the ideal MHD relaxation, and refer to $t=0$ as the start of this ideal MHD evolution. Unless otherwise stated all data and plots refer to the highest resolution $480^2\times 240$ simulations. % for the corresponding $k$ value.
\subsection{Qualitative description of the evolution}
%%%%%%%%%%%%
\begin{figure*}
\centering
%\hspace{0.2cm}
%(a)\includegraphics[width=3.5cm]{jiso60_p5x.eps}\hspace{1.2cm}
%(b)\includegraphics[width=3.5cm]{jiso60_p6w.eps}\hspace{1.2cm}
%(c)\includegraphics[width=3.5cm]{jiso60_p7w.eps}\\
%\includegraphics[width=5cm]{jcont_p5x.eps}
%\includegraphics[width=5cm]{jcont_p6w_l1.eps}
%\includegraphics[width=5cm]{jcont_p7w.eps}\\
\includegraphics[width=15cm]{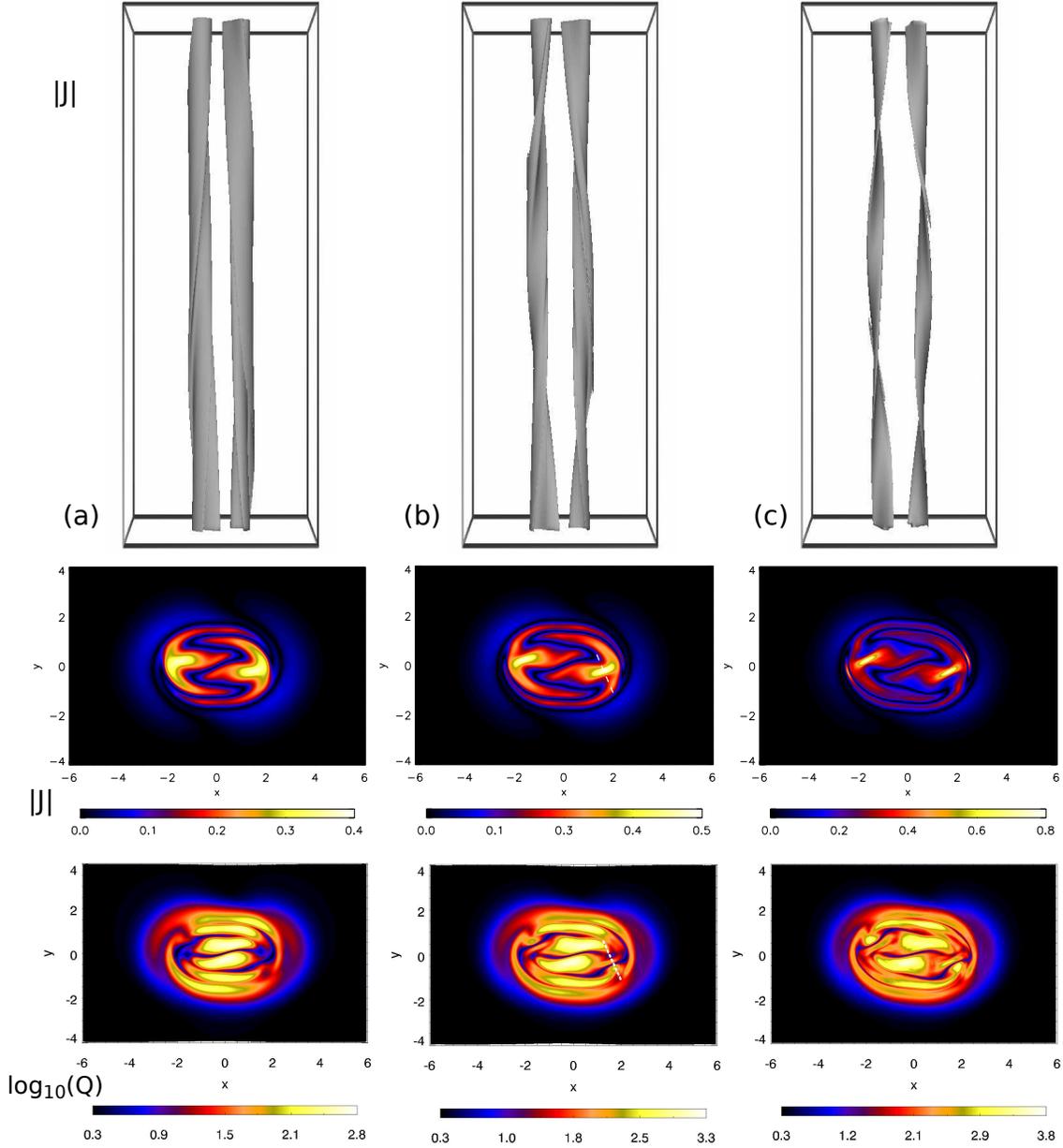}
\caption{Above: isosurface of the current density modulus $|\JJ|$ at 60\% of maximum. Middle: $|\JJ|$ in the plane $z=0$. {Below: $\log_{10}(Q)$ in the plane $z=0$}. For (a) $k=0.5$, (b) $k=0.6$, (c) $k=0.7$. Plots correspond in each case to the time when the measured current sheet thickness reaches its minimum value.}
\label{jstruc}
\end{figure*}
%%%%%%%%%%%%
%%%%%%%%%%%%
\begin{figure}
\centering
\includegraphics[width=6.5cm]{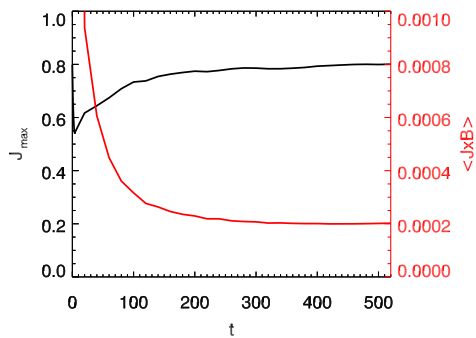}
\caption{Maximum current density $|\JJ|$ and domain mean of the Lorentz force $<\!\JJ\times\BB\!>$, for the run with $k=0.7$.}
\label{j_lor_ev}
\end{figure}
%%%%%%%%%%%%
The qualitative behaviour during the ideal relaxation for $0.5\leq k \leq 0.75$ can be described as follows. As the ideal MHD evolution proceeds, the current in the domain concentrates into two `ribbon' structures that run over the length of the loop in $z$, see Fig.~\ref{jstruc}. This is accompanied by a gradual growth (modest for small $k$) of the peak current in the domain. Eventually, after a period of between 5 and 20 Alfv{\' e}n transit times along the loop, the peak current plateaus  when an approximate equilibrium is reached. This is illustrated  in Fig.~\ref{j_lor_ev} where the maximum current density  and average $\JJ\times\BB$ force in the domain are shown for one representative value of $k$ ({note that the sharp drop in $J_{max}$ from $t=0$ to $t=1$ is due to fluctuations introduced by the interpolation between grids used by the two different codes, but these are quickly equilibrated}). The finite value that $<\!\JJ\times\BB\!>$ approaches is due to the finite gas pressure -- however, the field is as close to force-free as one would expect at coronal levels of the plasma-$\beta$ (note: $|\BB|, |\JJ|_{max}\sim\mathcal{O}(1)$). For comparison, the maximum $\JJ\times\BB$ force, excluding the region immediately next to the line-tied boundaries, is ~$2\times10^{-3}$ for $k=0.5$ and ~$2\times 10^{-2}$ for $k=0.75$.

As shown in Fig.~\ref{jstruc}, for higher values of $k$, the current density is localised into progressively thinner ribbons, attaining increasingly high peak values. For values $k\geq 0.8$ we are unable to reach an equilibrium because we are unable to sufficiently resolve the current layers, as discussed below.

%%%%%%%%%%%%%%%%
\subsection{Structure of current layers}
Examining the lower plots in Fig.~\ref{jstruc}, it appears that the final equilibrium of the ideal relaxation contains current ribbons that are of a finite thickness. In order to verify this, we repeat our simulations at a series of different numerical resolutions, and check for convergence of the current layer thickness with increasing resolution. More specifically, if there is some underlying, unresolved current sheet (possibly singular) then one would expect the measured current layer thickness to decrease proportional to the decreasing grid spacing. However, for the range $0.5\leq k \leq 0.75$ we observe a convergence of the current layer thickness with increasing resolution, indicating that we have a well-resolved, finite current layer. 
%%%%%%%%%%%%
\begin{figure}
\centering
(a)\includegraphics[height=6cm]{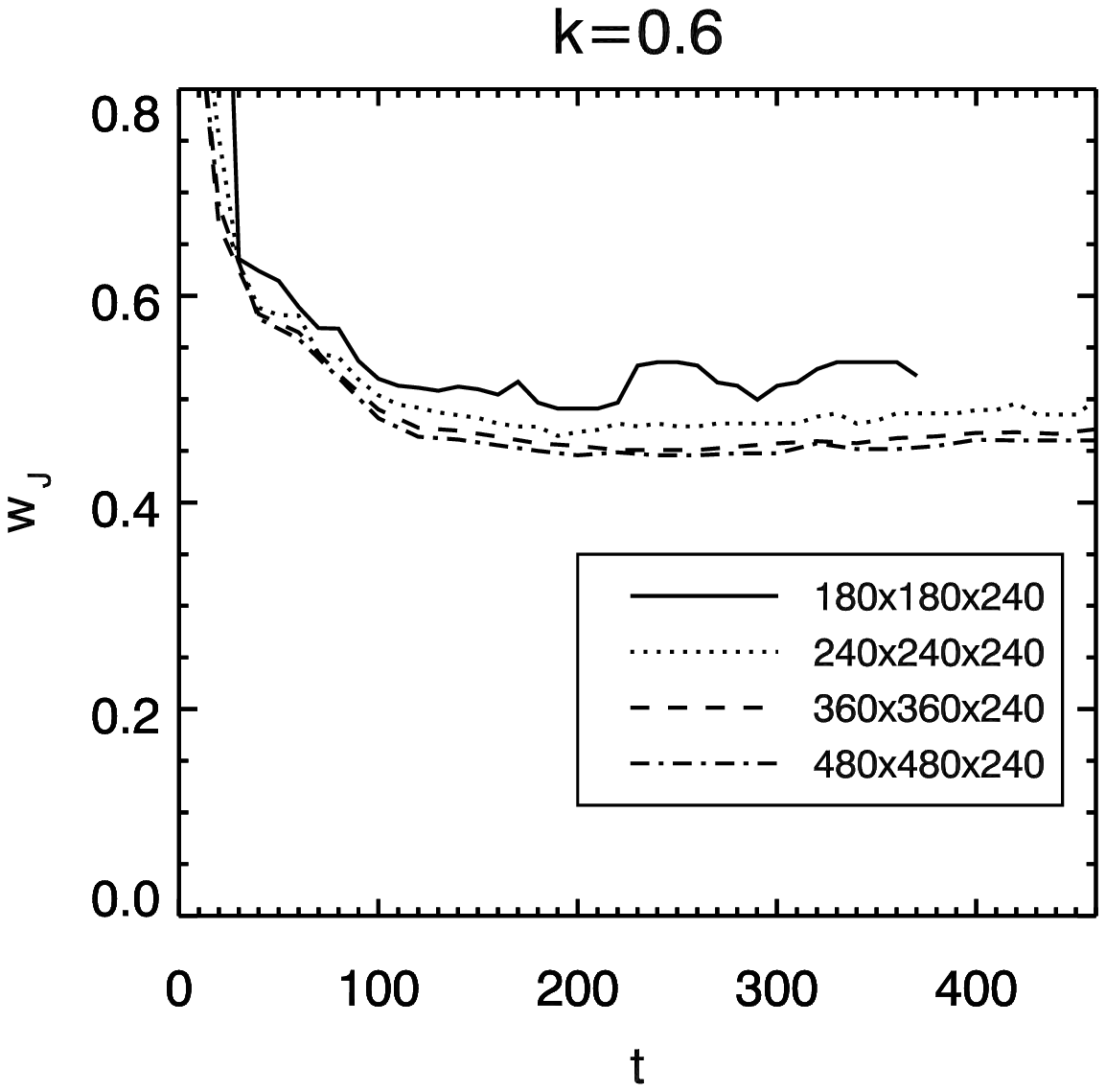}
(b)\includegraphics[height=6cm]{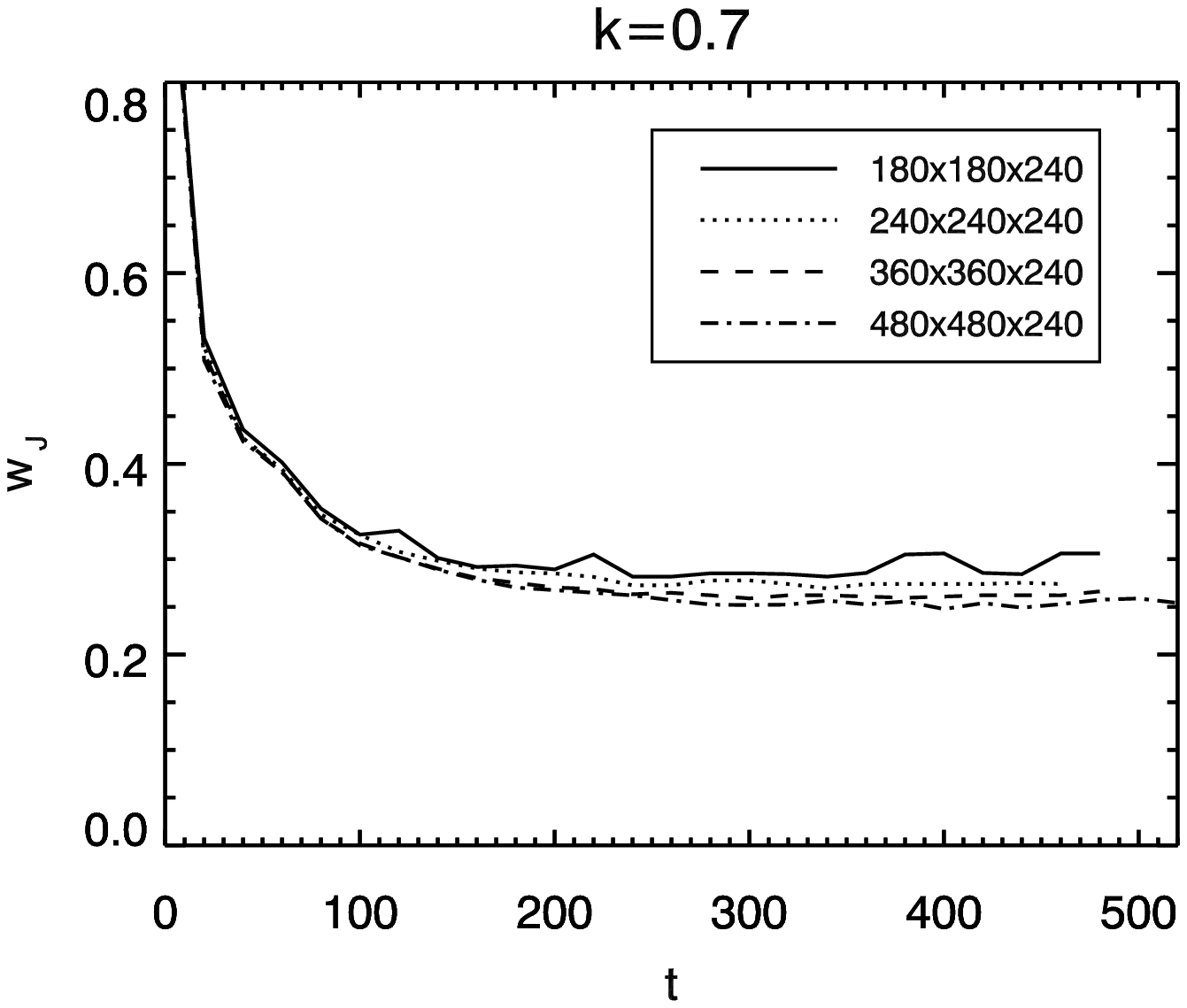}
\caption{Thickness of the current ribbon, $w_{J}$, measured in the $z=0$ plane as a function of time, for simulation runs with resolution $180^2\times240$ (solid line), $240^3$ (dotted), $360^2\times 240$ (dashed), $480^2\times 240$ (dot-dashed), for (a) $k=0.6$, and (b) $k=0.7$.}
\label{minwidth_p6p7}
\end{figure}
%%%%%%%%%%%%
This can be seen in Fig.~\ref{minwidth_p6p7}, where the thickness of the current layer for simulations with different resolution for $k=0.6$ and $k=0.7$ is shown. This layer thickness is defined as the full width at half maximum (f.w.h.m.) taken along a line through the current maximum, perpendicular to the layer -- see for example the overlayed dashed line on Fig.~\ref{jstruc}(b). 
{Generally speaking, for higher numerical resolution the small but finite numerical dissipation of the current layers decreases -- Fig.~\ref{minwidth_p6p7} shows that the final current layer thickness tends towards a fixed limiting value as the resolution is increased. As demonstrated in Appendix \ref{appA} this numerical dissipation has a negligible effect on the field topology.}
The peak current and current layer thickness exhibit only a very weak dependence on $z$. The current ribbons are found to have a minimum thickness somewhere within $-2<z<2$, and so for consistency we measure the thickness always in the $z=0$ plane. 

\begin{figure}
\centering
(a)\includegraphics[width=4cm]{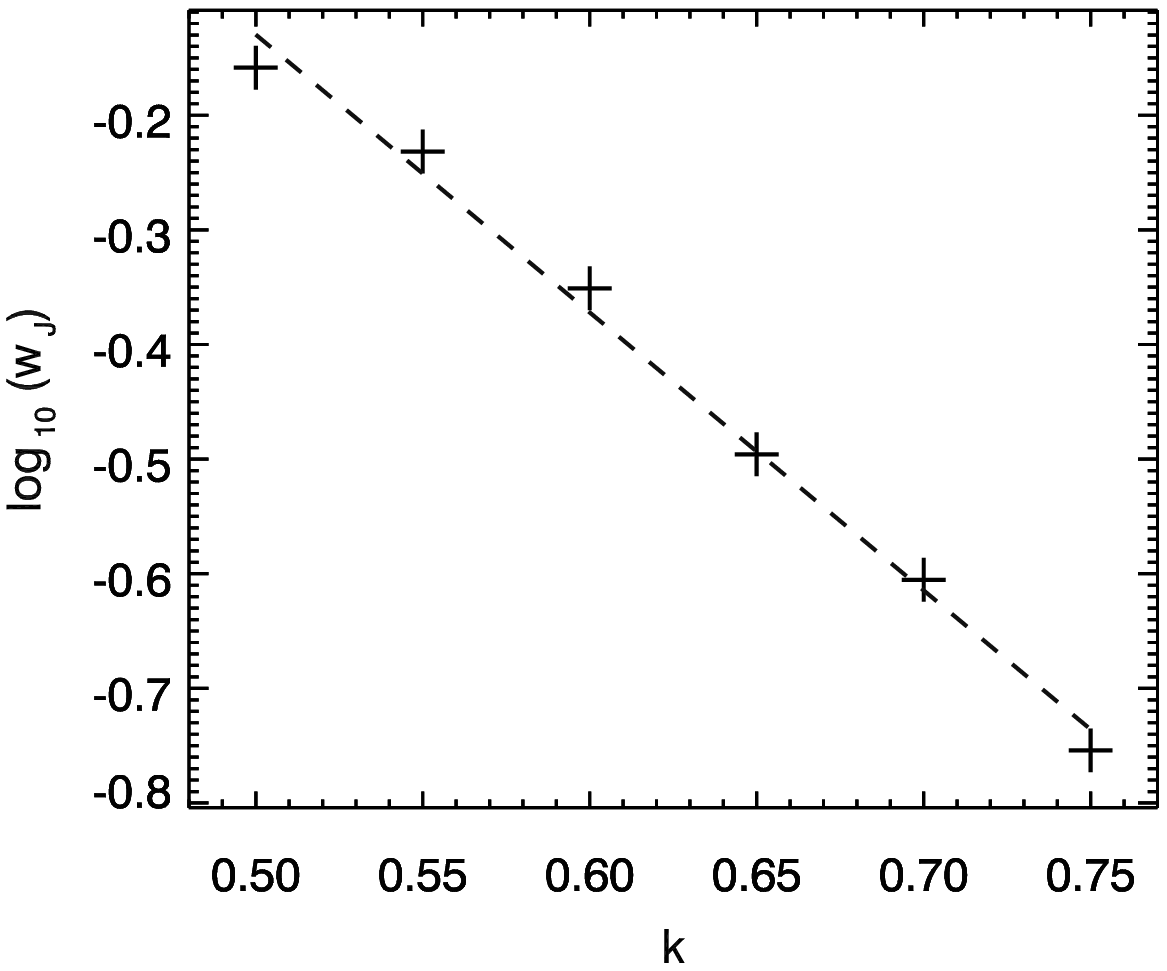}
(b)\includegraphics[width=4cm]{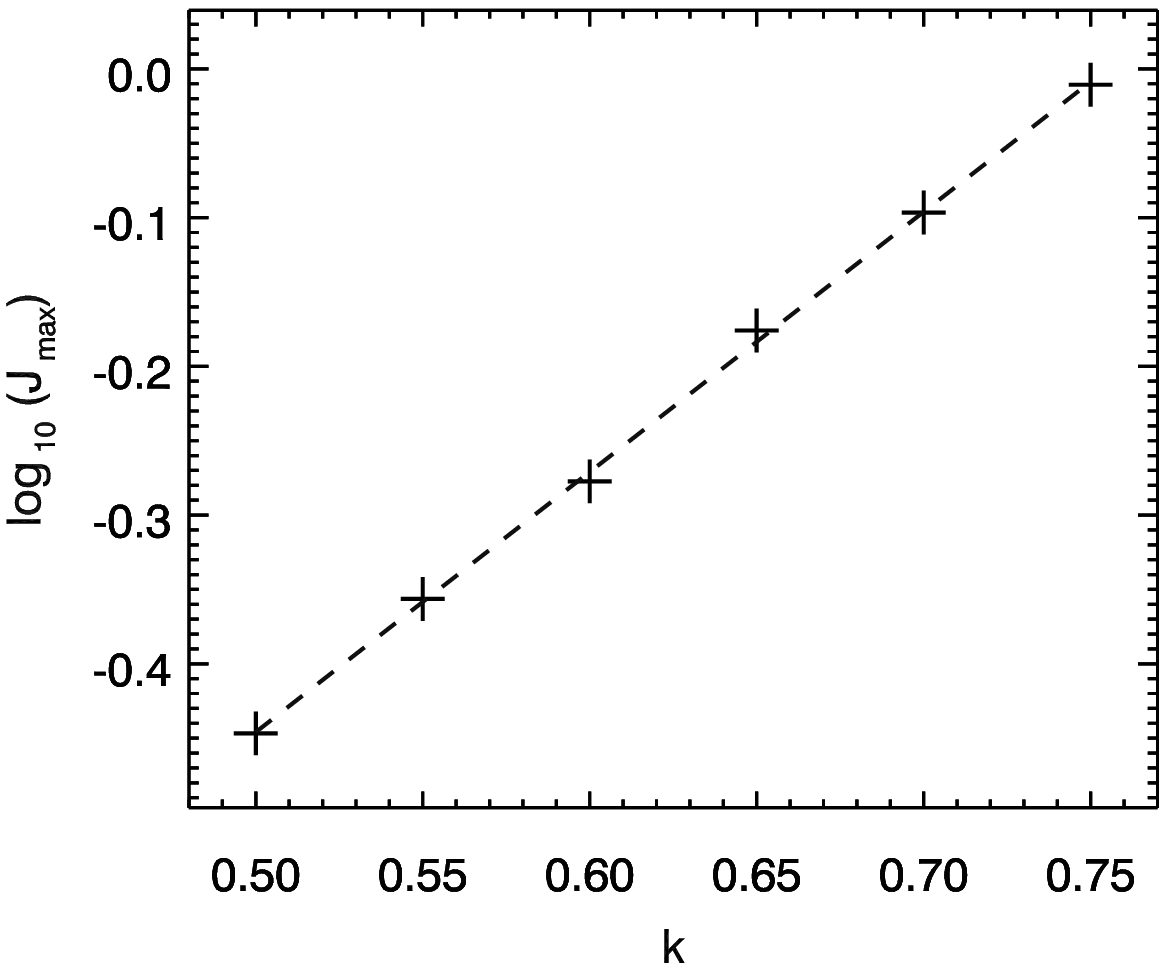}
\caption{(a) Thickness of the current ribbon in the relaxed state as a function of $k$. The dashed line represents~ $\log_{10}\, w_{J} =  -2.43k+1.10$. (b) Maximum current density in the domain as a function of $k$. The dashed line is~ $\log_{10}\, J_{max} =  1.97k -1.44$.}
\label{jscale}
\end{figure}

We now analyse the scaling of the current layer thickness in the relaxed equilibrium as a function of $k$ {(for a discussion on how this parameter can be related to observable quantities see Section \ref{obsec}.)} We take the current layer thickness to be the minimum thickness measured in the highest resolution simulations. This is expected to over-estimate the layer thickness due to the problem of defining the shortest path across the non-planar current layer in the automated procedure, but as shown in Fig.~\ref{minwidth_p6p7} the measured values are rather robust over time. The minimum current layer thickness $w_{J}$ as a function of $k$ is plotted in Fig.~\ref{jscale}(a). 
%%%%%%%%%%%%
%%%%%%%%%%%%
An exponential scaling is observed, following
\begin{equation}\label{minwidtheq}
\log_{10}\, w_{J} =  k (-2.43 \pm 0.11)+1.10 \pm 0.08.
\end{equation}
The peak current obtained in the simulations also depends exponentially on $k$, according to
\begin{equation}\label{jmaxeq}
\log_{10}\, J_{max} =  k  (1.97 \pm 0.09) -1.44 \pm 0.07.
\end{equation}
Uncertainties in the individual measurements of $w_{J}$ and $J_{max}$ are difficult to determine. As such, the error estimates given are 1-sigma error estimates, where a multiplying factor of $\sqrt{\chi^2/(N-M)}$ is included as described in \cite{bevington2003}, $N-M$ being the number of degrees of freedom (fits performed using the \verb poly_fit ~package in IDL). {Eliminating $k$ between Eqs.~(\ref{minwidtheq}) and (\ref{jmaxeq}) we have in addition that}
\begin{equation}
\log_{10} w_J =  (\log_{10} J_{max})  (-1.23 \pm 0.08) -0.68 \pm 0.16.
\end{equation}

These scalings can help us understand why we are unable to find numerically a force-free equilibrium for the case with $k=1$ studied by, e.g.~\cite{wilmotsmith2009a,wilmotsmith2010}. According to this scaling we expect a current ribbon with thickness $w_J(k\!=\!1)\approx 0.047$. At grid resolution $480^2\times240$ we have $\Delta x =0.025, \Delta y=0.017$, i.e.~the current layer thickness should be on the order of the grid scale. In order to properly resolve the current layer with minimal numerical diffusion we probably require at least 10-15 points across the current layer, so to find the equilibrium for $k=1$ we would need to increase the resolution in the $xy$-plane by a factor of 5-10, which is not possible within current computational limitations.

%%%%%%%%%%%%%%%%
\subsection{Magnetic energy}\label{efree_sec}
\begin{figure}
\centering
\includegraphics[width=5cm]{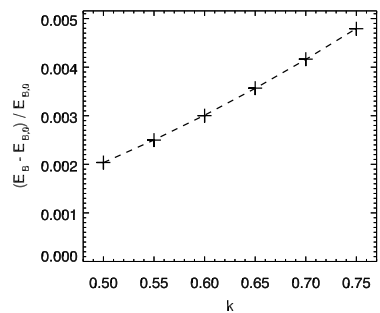}
%(b)\includegraphics[width=4cm]{scale_bxy.pdf}
\caption{Relative magnetic energy in the final state of the simulations as a fraction of the energy of the background field (see Eq.~(\ref{efree_def})), as a function of $k$.}% (b) Maximum value of $B_\perp$ in the domain in the final state as a function of $k$.}
\label{efree}
\end{figure}

We now calculate the magnetic energy above that of the homogeneous background field -- with a view to discussing the implications in Section \ref{efreedisc}. Specifically, we discuss here the relative energy of $\BB$ with respect to the background field, i.e. 
\begin{equation}\label{efree_def}
{E_{\rm rel}}=(E_B-E_{B,0})/E_{B,0}
\end{equation}
where $E_B=\int B^2/2\, dV$ and $E_{B,0}$ is the energy of the homogeneous background ($\BB=B_0\ee_z$) over the approximate domain of braided flux, a cylinder of radius 3 -- see Fig.~\ref{qmaps}. (Note that the code used is non-dimensionalised by setting the magnetic permeability $\mu_0=1$.) 
$E_{\rm rel}$  for the magnetic field of Eq.~(\ref{beq}) is given by $E_{\rm rel}=3.5\times 10^{-2}k^2$. That is, the total energy for  $0.5\leq k\leq 0.75$ is only around $1$--$2\%$ greater than the energy of the background homogeneous field. Around three quarters of this energy is removed during the magneto-frictional relaxation {(via frictional damping)}. The magnetic energy decays further during the ideal MHD evolution {(via viscous damping)}, to a level still above that of the homogeneous background. In order to liberate any additional magnetic energy requires reconnection to facilitate a simplification of the magnetic topology. This is prohibited here, and we are left with some non-zero final magnetic energy, which we associate with the minimum magnetic energy for the given topology. As expected, this energy is larger for larger values of $k$, and is plotted as a function of $k$ in Fig.~\ref{efree}(a). Since the energy of the field (\ref{beq}) depends quadratically on $k$ we fit this data with a quadratic in $k$, obtaining
\begin{eqnarray}\label{efree_eq}
E_{\rm rel}&=&10^{-3}\left[\right. k^2(8.60\pm 0.51) \nonumber\\
&&\left.+k(0.33\pm0.64) -0.27\pm 0.20  \right].
\end{eqnarray}
%%%%%%%%%%%%
%%%%%%%%%%%%

%%%%%%%%%%%%%%%%
%%%%%%%%%%%%%%%%
\section{Discussion}\label{discusssec}
\subsection{Relating current layer thickness and field line mapping: simulation results}\label{maplayers}
We have seen that the current layers in the relaxed state become increasingly thin as the complexity of the field is increased. However, given the restrictions of numerical resolution, these current layers are still so thick that one would not expect significant reconnection at coronal parameters. We would therefore like to use our results to extrapolate to larger values of $k$, in order to determine how much braiding is required to obtain sufficiently thin current sheets that one would expect the onset of magnetic reconnection in the corona, triggering a turbulent cascade that converts magnetic energy to thermal energy.

We first make more concrete the link between the field line mapping complexity and the current layers in the relaxed state. 
In order to do this, we use the $Q$ maps as quantifications of this complexity, and  measure the dimensions of the thinnest QSLs present. {For large values of $k$ a prohibitively high resolution field line grid would be required to properly resolve the high-$Q$ layers. We thus} take six representative cuts across the $Q$-maps, along $x=0,0.6,1.2$ and $y=0,0.55,0.1$, as shown by the black dashed lines overlayed on Fig.~\ref{qmaps}(c). We seek the peak in $Q$ which is thinnest over all six cuts, where in order to be counted the peak should have $Q$ at least 25\% of the domain maximum (in order to exclude very thin but weak layers). {In order to reach a situation in which the QSLs are resolved, the number of field line footpoints along the cuts is successively doubled until the peak value of $Q$ and minimum QSL thickness saturate (for $k=1$ this requires $2.4\times 10^4$ field line footpoints along the direction of the cut). For a discussion on such convergence procedures for $Q$ see \cite{aulanier2005}.}

\begin{figure}
\centering
\includegraphics[width=4cm]{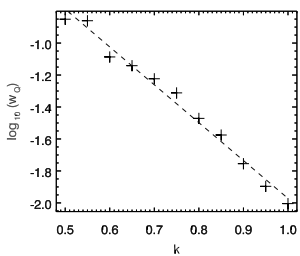}
\caption{Minimum thickness of layers in the squashing factor $Q$ representing the field line mapping, as a function of $k$.}
\label{qscale}
\end{figure}

The resulting minimum $Q$ layer thickness, $w_{Q}$, is plotted as a function of $k$ in Fig.~\ref{qscale}.
%%%%%%%%%%%%
%%%%%%%%%%%%
%%%%%%%%%%%%
\begin{figure}
\centering
\includegraphics[width=6.5cm]{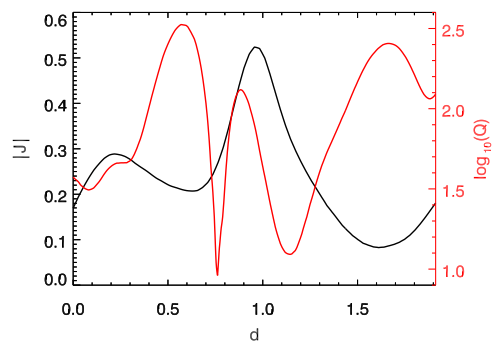}
\caption{$|\JJ|$ (black) and $Q$ (red) along the dashed line shown in Figure \ref{jstruc}(b), for the run with $k=0.6$.}
\label{j_q_ajl}
\end{figure}
%%%%%%%%%%%%
The thickness of these $Q$ layers is seen to decrease exponentially as $k$ is increased, with the best fit line on the plot corresponding to a scaling of 
\begin{equation}\label{qscaleeq}
\log_{10} (w_{Q}) =  k (-2.37 \pm 0.09) + 0.40 \pm 0.07.
\end{equation}
Comparing this with Eq.~(\ref{minwidtheq}), we see that within the error bars, the thickness of the $Q$ layers and the thickness of the current layers in the relaxed state follow an identical scaling. This provides strong evidence of a causal link, i.e. current layers in the force-free equilibrium form on the scale of the layers present in the field line mapping. Interestingly, the braided field in the centre of the domain is characterised by a 3D volume within which $Q$ is large, containing many layers wrapped around one another. 

The association of isolated QSLs with current build-up is {well documented \cite[e.g.][]{demoulin96a,demoulin1997,demoulin2006}. Recent analyses show excellent correlation between $Q$ and $\JJ$ in simulations and observations of solar flares in certain generic configurations \citep[e.g.][]{savcheva2012,janvier2013}.}
{However, comparing directly the profile of $Q$ and the current density in our simulations, we observe no direct spatial correlation between their maxima. From the middle and lower images in Figure \ref{jstruc} we see that the locations of the current layers do not correspond to the field lines with highest values of $Q$ -- however, there is a clear enhancement in $Q$ around the current layer for $k=0.6,0.7$. Figure \ref{j_q_ajl} shows the two quantities along a cut through the current layer (dashed line in Figure \ref{jstruc}b) for $k=0.6$. Again, we see a local maximum of $Q$ close to the local maximum of $|{\bf J}|$, but the two curves display different profiles, and there are adjacent higher maxima of $Q$. (Note that the grid spacing in the $xy$-plane at resolution $480^2\times 240$ means that the scales in both quantities are well resolved.)}

%What we show here is that even in the case where the high-$Q$ region has a volume-filling effect --  forming a volume-filling foliation of QSLs rather than a single isolated QSL -- the thickness of the $Q$ layers is still integral to the formation of the current layers, {\bf even though there is no direct link between locations of maximum current and maximum squashing factor}. 
{Despite the fact that there is no direct link between locations of maximum current and maximum squashing factor, we find here  that the thickness of the $Q$ layers is still integral to the formation of the current layers, and that this is the case
even when the high-$Q$ region has a volume-filling effect --  forming a volume-filling foliation of QSLs rather than a single isolated QSL}. 
Evaluating the scaling with $k$ of the maximum value of $Q$ over the whole domain, we obtain
\begin{equation}\label{qmaxscaleeq}
\log_{10} (Q_{max}) =  k ( 4.70 \pm 0.20) +  0.80 \pm 0.15.
\end{equation}
{Eliminating $k$ between Eqs.~(\ref{qscaleeq}) and (\ref{qmaxscaleeq}) we have in addition that}
\begin{equation}\label{combinedqscale}
\log_{10} w_Q =  (\log_{10} Q_{max})  (-0.50 \pm 0.03) +0.80 \pm 0.11.
\end{equation}
%\tg{see also Figure \ref{qscale}(b)}.
{The subtle relation between $Q$ and the current layer thickness will be explained in the next section.}
%We therefore conclude that it is not the value of $Q$, but the thickness of the scales in $Q$ (i.e.~of the QSLs) that provide a prediction of the nature of the final state. 

%%%%%%%%%%%%%%%%%
\subsection{Relating current layer thickness and field line mapping: theoretical considerations}\label{maplayers:theory}
{In order to understand why the current layers form on the same scale as layers in the field line mapping, consider that the final equilibrium satisfies
\begin{equation}
\nabla\times\BB\approx \alpha\BB.
\end{equation}
In a perfect equilibrium, $\alpha$ is constant along field lines. Due to the complex field line mapping, $\alpha$ must vary rapidly in the direction perpendicular to $\BB$, as the following demonstrates. Eq.~(\ref{qeq}) can be written in the form
\begin{equation}
Q=||DF||^2/\det(DF),
\end{equation}
where $F=(X(x,y),Y(x,y))$ is the field line mapping from $z=-24$ to $z=24$ and $DF$ is its Jacobian. Now let the eigenvalues of $DF$ be $\lambda_{max},\lambda_{min}$ (which are both positive since there are no nulls in the domain). Then by definition an elemental flux tube is stretched by a factor $\lambda_{max}$ in one direction, and squeezed by a factor $\lambda_{min}$ in the orthogonal direction, generating small scales in the mapping if $\lambda_{max}\gg\lambda_{min}$ \citep{titov2002}.
Now suppose that $\alpha$ has some smooth distribution with typical scales $\ell$ on, say, $z=-24$. Then, since $\alpha$ is constant along field lines, we can map along the field lines to the upper boundary to find $\alpha(z=24)=\alpha(F^{-1}(x,y))$. The smallest length scales of $\alpha$ at $z=24$ are therefore of order $\lambda_{min}\ell$.}
{Furthermore, since $\alpha=\JJ\cdot\BB/B^2$, and since $|\BB|\approx 1$ here, we see that $J_\|$ is directly proportional to $\alpha$. We therefore conclude that $\JJ=J_\|\ee_B$ should also have scales of order $\lambda_{min}\ell$.} Here we associate $\ell$ with the length scale of the initial current distribution, $\ell\sim 1$, which is independent of $k$. 

{To relate the above length scales to $Q$ we note that since $B_z\approx 1$ at $z=\pm 24$, we have 
\begin{equation}\label{evaleq}
\det(DF)=\frac{{B_z}^-}{{B_z}^+}=\lambda_{max}\lambda_{min}\approx 1,
\end{equation}
where ${B_z}^-=B_z(z\!=\!-24)$ and ${B_z}^+=B_z(z\!=\!24)$
(to be precise, we have $0.97\leq \lambda_{max}\lambda_{min}\leq 1.03$.) Therefore
\begin{equation}
Q\approx||DF||^2,
\end{equation}
and in addition
\begin{equation}\label{rateq}
\frac{\lambda_{max}}{\lambda_{min}}=\frac{Q}{2}+\sqrt{\frac{Q^2}{4}-1}~~\approx Q
\end{equation}
for $Q\gg 1$ \citep[see][]{titov2002}. Thus, combining Eqs.~(\ref{evaleq},\ref{rateq}), 
\begin{equation}\label{evq}
\lambda_{min}\approx Q^{-1/2} ~~\Leftrightarrow~~ \log \lambda_{min}\approx -\frac{1}{2}\log Q.
\end{equation}
Examining Eq.~(\ref{combinedqscale}), we find this relation holds to a good approximation, certainly within the range of $k$ values considered.}

Note that the above arguments rely on $\alpha$ being constant along field lines. Since we do not obtain an exact force-free equilibrium, this is not exactly the case. Examining $\alpha^\star=\JJ\cdot\BB/B^2$ along field lines in the final state of the simulations, we find that for the field lines with largest average values of $\alpha^\star$, the variation of $\alpha^\star$ is of order $5\%$ for $k=0.5$ and of order $20\%$ for $k=0.75$. This can allow for some corresponding variation in $J_\|$ along field lines, so one might expect that the current scaling with $k$ could be a little weaker in the simulation results. This does not seem to be the case here. 
One should note that the argument above is developed for a single elemental flux tube, while in our field there is a continuous range of values of $Q$, and corresponding $Q$-layer thickness. We note in particular that the measured values of $Q_{max}$ and $w_Q$ typically do not occur on the same field line, i.e.~the thinnest layers are not necessarily associated with the highest values of $Q$. 
%This is consistent with the the observation of \cite{wilmotsmith2009b} that the current layers in the braided field do not form along field lines corresponding to the highest values of $Q$ in the initial state.

{The above argument demonstrates that for a force-free, braided magnetic field, we expect current layers that are (at least) as thin as small scales in the field line mapping, as determined by the smallest eigenvalue of the field line mapping, $\lambda_{min}$. In our simple case this is related to the value of $Q$ by  the relationship in Eq.~(\ref{evq}). This  can be generalised to the case in which the normal component of $\BB$ is not uniform on the boundaries. In that case, we can combine Eqs.~(\ref{evaleq},\ref{rateq}) to find
\begin{equation}
\lambda_{min}\approx\left(Q\frac{{B_z}^+}{{B_z}^-}\right)^{-1/2}.
\end{equation}}

It is expected that on the Sun as the magnetic field is continually tangled by the photospheric motions, the scales in the field line mapping get smaller at an exponential rate. This was predicted by the simulations of \cite{vanballegooijen1988a,vanballegooijen1988b,mikic1989}. It has also been confirmed by \cite{yeates2012} using measurements of photospheric flows derived from magnetograms taken by Hinode/SOT. As such, we expect that at some threshold level of braiding the high $Q$ layers -- and by extension the current layers using the above argument -- will become sufficiently thin that reconnection should occur. In Section \ref{onsetsec} we attempt to place an upper bound on this threshold level of braiding.

%%%%%%%%%
\subsection{Relation to observable quantities in solar observations}\label{obsec}
Our results support the hypothesis that turbulent photospheric motions naturally lead to the formation of thin current layers in the corona, and therefore heating. This is demonstrated here for one particular braiding pattern parameterised by parameter $k$. However, as shown in the previous section the length scales in the field line mapping for any generic braiding pattern are the crucial quantity for determining the expected current layer thickness in the corresponding equilibrium field. In order to relate these results to solar observations requires careful analysis to determine the length scales present in the magnetic field line mapping in the corona. One way to access this is to analyse the properties of magnetic field extrapolations, although with present resolution this is unlikely to reveal the smallest scales present. An alternative, highly promising, approach was employed by \cite{yeates2012}, who used local correlation tracking to determine a time-sequence of photospheric flows, and from that reconstruct the field line mapping in the coronal volume assuming an ideal evolution. They showed that measures of the field line mapping such as $Q$ and the finite-time Lyapunov exponent increased in net value and developed progressively smaller scales with time. What is required now is a systematic study in which this approach is extended to cover different observed regions and to include the structuring of the initial coronal field (\cite{yeates2012} assumed an initially homogeneous field). This requires high temporal and spatial resolution observations of photospheric magnetograms and thus flows. Such a study should shed light on the braiding timescale in different regions. In the following section we discuss when reconnection onset would be expected to occur as the braiding proceeds and length scales in the field line mapping become ever thinner.

%%%%%%%%%
\subsection{Onset of turbulent relaxation in the corona}\label{onsetsec}
\cite{pontin2011a} performed a series of resistive MHD simulations with increasing Lundquist numbers (based on the horizontal scale of the initial field) of $10^{2}-10^{4}$, for the field (\ref{beq}) with $k=1$. They showed that reconnection at the two initially formed current layers is followed by a turbulent cascade involving the formation of myriad current layers, through which the magnetic field `unbraids' itself, releasing stored magnetic energy and heating the plasma. However, in order for this turbulent cascade to be triggered one requires appreciable reconnection to occur in the two initial current layers, at coronal parameters. We cannot make a definitive statement here about when this would be expected: there is no well defined onset criterion for reconnection in three dimensions. Furthermore, estimating just how fast the flux must be reconnected in the first event is not straightforward. Below we make some estimates for the onset of reconnection in the current ribbons observed in our final equilibrium using established theories based primarily on two-dimensional geometries. We consider these to be conservative estimates for the triggering of the turbulent cascade, i.e.~the reconnection onset in this 3D geometry is likely to be sooner than these estimates predict.

One scenario in which fast reconnection can be triggered in the corona (in the presence of a strong guide field as relevant here) is when the current sheet thickness falls below the ion Larmor radius $r_L=c_s/\Omega_{ci}$, where $c_s=\sqrt{k_BT/m_i}$ is the ion sound speed, $\Omega_{ci}=eB/m_i$ is the ion cyclotron frequency and $T$ is the total temperature. This was demonstrated using numerical simulations by \cite{cassak2007} (see also \cite{uzdensky2007}) and observed in the laboratory by \cite{egedal2007}. Taking a typical coronal temperature of $10^5-10^6K$ and typical magnetic field strength of $10-100G$, then $r_L$ is approximately in the range $3-100cm$. We can now use Eq.~(\ref{minwidtheq}) to estimate the value of $k$ that would lead to current layers of this thickness in the relaxed state, associating length units in Eq.~(\ref{minwidtheq}) with $Mm$. Doing this, we obtain an estimate of $k\approx 2.9-3.5$. It should be noted however that it may not be necessary for the braiding complexity to be this high, for a number of reasons. First, the initial reconnection event need not be very fast; it is only necessary to reconnect a sufficient amount of flux on an Alfv{\' e}nic timescale. Once the turbulent relaxation is triggered, this can become globally fast through reconnection in myriad current layers as observed by \cite{pontin2011a}. Second, it was pointed out by \cite{cassak2006} that the reconnection process itself tends to drive current sheet thinning even in the absence of external forcing. Therefore slow Sweet-Parker-type reconnection in a moderately thin current ribbon (corresponding here to $k<2.9$) could itself drive the sheet thinning, rather than a further increase in the braid complexity (here, increase in $k$) being required.

Another known trigger for the onset of fast reconnection is the non-linear tearing (or `plasmoid') instability \citep{loureiro2007,bhattacharjee2009}. This is known to enhance the reconnection rate in both 2D and 3D current layers \citep{daughton2011,baalrud2012,wyper2014a}. The non-linear tearing instability typically requires a current sheet aspect ratio (in the plane of the magnetic shear -- here $xy$) of $50-100$. Since the current sheet cross-section is approximately independent of $z$ (and thinnest around $z=0$) we use the aspect ratio in this cross-section to make an estimate of when (for increasing $k$) we would expect the layer to undergo non-linear tearing. We evaluate the f.w.h.m.~both along and across the layer in this plane, and define the aspect ratio as the ratio of these two quantities. It turns out that both quantities decrease with increasing $k$, such that the aspect ratio (a.r.) as a function of $k$ follows an approximate linear relation given by 
\begin{equation}
{\rm a.r.}=k(15.4\pm1.1) -5.75\pm 0.69.
\end{equation}
Thus a critical aspect ratio of $50-100$ requires $k=2.9-6.1$. Again, for the reasons given above we expect that this is a conservative upper bound on the required value of $k$.

Finally, one can consider when non-MHD effects might become important due to the high electric field. In particular, electron runaway may occur when the electric field exceeds the Dreicer field \citep[e.g.][]{holman1985} -- it has been argued that for coronal parameters a non-classical dissipation in the layer is required to limit the particle velocity/acceleration length \citep[e.g.][]{martens1988,craig2002}, which may be orders of magnitude bigger than the Spitzer resistivity. A localised enhancement in the dissipation is known to be favourable for triggering fast reconnection. 
Taking $n=10^{10}cm^{-3}$ and $T=10^6K$ the Dreicer field in the corona is of order $10^{-2}Vm^{-1}$. Assuming a collisional Spitzer resistivity in the current layer one can translate the electric field threshold into a current threshold. Using the peak current scaling obtained in Eq.~(\ref{jmaxeq}), one can then relate this to the corresponding value of $k$.
Associating length in our simulation with $Mm$ as before, and taking $B_0=100G$, then the measured value of $|{\bf J}|_{max}$ in non-dimensional code units should be $10^6$. This is estimated to occur based on Eq.~(\ref{jmaxeq}) for $k=3.8$ (or for $B_0=10G$, $k=4.3$).

Summarising the above results, we can say with some confidence that $k\approx 3$ provides a conservative upper bound on the permitted braid complexity at coronal parameters: beyond this degree of braiding, we expect significant reconnection to occur in the current layers leading to the triggering of a turbulent relaxation, most likely due either to a transition to fast collisionless reconnection or the onset of the plasmoid instability. We note that one could also achieve the same braid complexity through many twists of lower strength. We return to consider different braiding patterns in Section \ref{diffbraid}. Note also that we have implicitly assumed above that no other instability sets in for lower complexity. One could imagine, for example, that at some point the loop may be susceptible to an ideal instability similar to a kink instability \citep[e.g.][]{hood2009}. However, our loop has no axis of symmetry as in a classical kink mode, and the net twist is zero. The existence of an instability analogous to a kink for such a loop is unknown. While in our simplified model there are large regions of coherent twist for $k\approx 3$, we would expect in reality the braid pattern would be much less coherent.

One leading theory for the onset of reconnection in the corona is the idea that there exists a {\it critical angle} for magnetic field lines on opposite sides of the current sheet \citep{dahlburg2001,dahlburg2005}. For a mismatch in $\BB$ greater than the critical angle, the {\it secondary instability} can lead to fast dissipation of energy. However,  this instability can only take effect after the primary instability has modified the current sheet equilibrium into a state containing a series of aligned flux tubes. This primary tearing instability has a slow growth rate, and as pointed out by \cite{dahlburg2005}, one has to invoke an anomalously high resistivity within the current layer in order to have a growth time that is shorter than the timescale of current layer build-up by the stressing motions. In the absence of such anomalous resistivity, the primary instability is too slow, and the secondary instability does not have the chance to set in.
Of course, if non-linear tearing eventually provides the trigger for the instability (when the current sheet aspect ratio reaches some threshold, see above), then this secondary instability can still play a key role.

%%%%
\subsection{Free energy}\label{efreedisc}
One crucial parameter in any nanoflare heating model is the free energy stored in the magnetic field that is available for conversion into heat. We can obtain the energy prior to reconnection onset using Eq.~(\ref{efree_eq}). We also need to estimate how much of this energy is available to be released during the turbulent relaxation. In \cite{pontin2011a} it was shown that the asymptotic state of the turbulent relaxation for $k=1$ is not the homogeneous field, as predicted by the Taylor relaxation hypothesis, but rather that only around 60\% of $E_{\rm rel}$ was released before the field settled to a new non-linear force-free field (without current layers). Without any data for different $k$ values, we make the assumption that $E_{\rm free}(k) \approx 0.6 E_{\rm rel}(k)$ provides a representative estimate for the energy available for all $k$, where $E_{\rm free}$ is the maximum available nanoflare energy.

To arrive at a dimensional quantity  for the energy released, we analyse the volume of braided flux, approximately a cylinder of radius 3 in the simulations, and associate length units in our code with $Mm$, as above. This means our loop has radius $3Mm$ and length $48Mm$. Supposing that the energy release is triggered for $1\lesssim k\lesssim 3$, and setting $B_0=10-100G$, we find for $k=1$ that $3\times 10^{25}{\rm ergs}\leq E_{\rm free}\leq 3\times 10^{27}{\rm ergs}$, while for $k=3$ we predict $2\times10^{26}{\rm ergs}\leq E_{\rm free}\leq 2\times10^{28}{\rm ergs}$. The lowest of these estimates is consistent with the upper range of nanoflare energies considered by \cite{cargill2014}, though the others are some orders of magnitude larger. Therefore the available energy in the braided field seems to be more than sufficient to provide nanoflare energy release in the desired range. Indeed, we do not expect this entire energy content to be released in a  single distinct event. It was shown by \cite{pontin2011a} that the timescale for turbulent relaxation is at least on the order of a few Alfv{\' e}n crossing times along the loop, and is expected to be longer for smaller {resistivity}. Therefore it is expected that at coronal parameters the field will never reach the minimum energy state, but will be continually driven away from this state by turbulent convective driving motions.
Eventually, the frequency of energy release events -- and overall heating rate -- will then be determined by a balance between the reconnection-mediated energy release that acts to untangle the field, and the photospheric driving that typically acts to increase the braid complexity. Such behaviour has been confirmed in ongoing studies with continual boundary driving \citep{ritchie2015}.

\subsection{Perpendicular magnetic field}
The free energy in the domain is associated with magnetic field components perpendicular to the tube axis, $B_\perp=|\BB_{xy}|$. 
Based on energy balance arguments \cite{parker1988} asserted that the average coronal value of this field component should be of the order 25\% of the axial field, giving the `Parker angle' of field lines of around $14^\circ$ to the vertical. 

Some theories propose that when $B_\perp$ reaches a critical value in the corona, reconnection is initiated. In particular, the {\it secondary instability} sets in approximately when field lines have a $20^\circ$ angle to the vertical, or when $B_\perp\approx 0.35 B_z$ \citep{dahlburg2005}.
\cite{berger2009} studied a discrete strand-based model for a braided loop and, citing the driven reconnecting flux tube simulations of \cite{linton2001}, imposed a critical angle between neighbouring strands of around $30^\circ$ (corresponding to $B_\perp\approx0.27B_z$). 
In our simulations the asymptotic value of the domain maximum of $B_\perp$ as a function of $k$ is found to follow the scaling
\begin{equation}
B_{\perp, max}= k(0.417\pm0.008) -0.060\pm0.005.
\end{equation}
Even for the simulation run with $k=0.75$ considered here, $B_\perp\approx0.26B_z$, yet at coronal parameters we would not expect significant reconnection to occur in this configuration. Extrapolating to $1\leq k\leq 3$ we expect $0.4B_z\leq B_\perp\leq 1.2 B_z$. Thus the magnetic shear across the reconnecting current layer may be significantly larger than previously anticipated.
Interestingly, this also calls into question the applicability of the reduced MHD approximation in addressing the topological dissipation hypothesis when braided field structures of the complexity of those studied here are considered, as this approximation requires $B_\perp\ll B_z$.

%%%%
\subsection{Relation to previous ideal relaxation results}
In general, for a given magnetic topology there may exist multiple force-free equilibria. We previously reported \citep{wilmotsmith2009a} that an ideal relaxation of a braided magnetic field (defined by Eq.~(\ref{beq}) with $k=1$) using a magnetofrictional (MF) approach led towards a force-free equilibrium with no small-scale current layers. In the present study we have used a different relaxation method and found a different equilibrium. The equilibrium we obtained here has a lower magnetic energy than that obtained by the MF approach. Nonetheless, we still find no evidence for formation of tangential discontinuities, but rather current layers of finite thickness are present.

The reason that we obtain here a lower energy state is likely because our relaxation follows a different path. The absence of inertia and strict monotonic decay of magnetic energy, $E_M$,  in the MF evolution mean that the relaxation is much more likely to halt at an intermediate local energy minimum. By contrast, in our ideal MHD simulations $E_M$ is first converted to kinetic energy, $E_K$, with this in turn being damped by viscosity. Thus, it is not $E_M$, but $E_M+E_K$ that is strictly decreasing. This additional freedom that is allowed in the ideal MHD evolution leads us to believe that the equilibria found here are probably global energy minima for the topologies considered.  
An alternative to the above interpretation is that the MF method may approach a {\it numerical} equilibrium that is too far from the equilibrium found here to show the formation of the current layers. The MF relaxation is known to be compromised when distortions of the computational mesh lead to local inaccuracies in the evaluation of the $\JJ\times\BB$ force \citep{pontin2009}, though these errors can be significantly reduced by implementation of a `mimetic' approach to calculate derivatives, as performed recently by \cite{candelaresi2014}. % Finally, it is possible that the small finite change of topology is the present simulations is important in finding the equilibrium. Note however that this change is minimal (see \ref{appA}), and what we can say for certain is that smooth equilibria containing thin current layers exist for a sequence of braided fields (the final states of our simulations)} 

Further analysis of the MF relaxation results revealed that transferring the final state to a resistive MHD code led to a rapid collapse of the current to small scales \citep{wilmotsmith2010}. This was previously thought to be a result of some instability, either ideal or resistive, of the associated energy minimum. However, it now seems likely that the perturbation associated with interpolating to a new grid was sufficient to dislodge the system from the local energy minimum it was in, allowing a further evolution towards the state containing thin current layers.
With the present results we see that it is not necessary to invoke an  instability to explain the formation of these current layers. Rather, the ideally accessible force-free equilibrium contains thin current layers, on the order of the grid spacing for $k=1$. Therefore even the numerical dissipation would be sufficient to induce reconnection, thus setting off the turbulent relaxation observed by \cite{pontin2011a}.

We should also note here how our results relate to those of \cite{vanballegooijen1988a,vanballegooijen1988b}. In those papers, a sequence of shear velocities was applied to the top and bottom perfectly conducting plates of a volume containing an initially uniform field. After each subsequent shear perturbation a corresponding equilibrium was found using an iterative energy minimisation approach. \cite{vanballegooijen1988a} noted an exponential decrease in the scales of the field line mapping (though this was before the development of the theory regarding the squashing factor), and predicted that the current in the relaxed state should form on a corresponding scale. \cite{vanballegooijen1988b} subsequently demonstrated numerically that current layers of finite thickness developed, and indeed exhibited an exponential decrease in their thickness with each successive displacement. Our results are entirely consistent with these, both in the exponential scaling of the QSL thickness with the perturbation and in the link between QSL thickness and current layer thickness.

\subsection{Scaling of $Q$ layers for different braided fields}\label{diffbraid}
We have examined here the formation of current layers for one particular category of braided magnetic fields, linking this to small scales present in the field line mapping. It is instructive to analyse briefly the differences one would expect for different braiding patterns. 

We first consider a braiding pattern with non-zero net twist, obtained by removing the factor $(-1)^i$ from Eq.~(\ref{beq}) so that all six `twist regions' have the same sign. As discussed in \cite{wilmotsmith2011b}, this field has a comparable maximum value of $Q$ to that above -- at least for $k=1$ -- but the field line mapping is significantly less complicated globally, i.e.~the region of high $Q$ fills the space much less effectively (see Figure 2 of \cite{wilmotsmith2011b}). Additionally, the field releases around half as much energy as the one studied above during a turbulent relaxation (again, for $k=1$). 
Now suppose that there is a direct relation again in this case between the thickness of current layers in the ideally relaxed state and the thickness of layers in the field line mapping (as expected due to the arguments presented in Section \ref{maplayers:theory}). Performing the same calculation as in Section \ref{maplayers} we can estimate the scaling of the $Q$ layer thickness with $k$. We find that
\begin{equation}\label{qscaleeq1}
\log_{10} (w_{Q}) =  k (-1.91 \pm 0.06) + 0.08 \pm 0.04.
\end{equation}
Now by the same arguments as put forward in Section \ref{onsetsec} we find that we require $k\approx 3.2-4.0$ to reach the threshold for collisionless reconnection onset. However, since the field is globally twisted, it is more likely in this case that some kink-type ideal instability will set in before this limit is reached.

\cite{wilmotsmith2009b} considered a different braiding pattern, again related to that in our original model. They fixed $k=1$ in Eq.~(\ref{beq}) and considered the scaling of the QSL thickness as the number of of twist pairs is increased -- parameter $n$ in Eq (\ref{beq}). They found that the QSL thickness scaled, again exponentially, as $1.90\times 10^{-0.93n}$. Therefore in order to reach a given thickness of QSL layer, one requires marginally less total twist (equating approximately, say, a doubling of $k$ with a doubling of $n$). This is expected on the grounds that the combination considered by \cite{wilmotsmith2009b} constitutes a maximally efficient protocol for generating small scales as measured by stretching of material lines under the mapping \citep{boyland2000}.

%%%%%%%%%%%%%%%%
%%%%%%%%%%%%%%%%
\section{Conclusions}\label{concsec}
In this paper we investigated the ideal relaxation of  a class of braided magnetic fields with non-trivial field line mapping. The key results, and implications of those results, are as follows.
\begin{itemize}
\item
Smooth force-free equilibria do exist for these braided fields. The equilibria contain thin but finite current layers, whose thickness decreases with increasing braid complexity. 
\item
Specifically, the current layer thickness is directly related (follows an identical scaling with our complexity parameter $k$) to the small length scales shown by the field line mapping. 
%, as measured by the squashing degree $Q$ (i.e.~the width of QSLs). We note however that the value of $Q$ itself does not follow a scaling directly related to the current layer thickness or the peak current, nor does its maximum provide a prediction of the exact formation location of the current layers.
\item
{This result can be generalised (see Section \ref{maplayers:theory}): for any braided magnetic field (in which $\BB\neq{\bf 0}$), if a force-free equilibrium exists then it should contain current sheets at least as thin as the smallest length scales present in the field line mapping, as determined by the smallest eigenvalue of the field line mapping, $\lambda_{min}$.}
\item
The thickness and intensity of the current layers obtained after the relaxation both scale exponentially with the parameter of braiding complexity, $k$. This is consistent with previous works who described an exponential scaling with number of boundary shears \citep{vanballegooijen1988a,vanballegooijen1988b,mikic1989}, and demonstrates that under continuous driving a coronal loop will quickly reach a state in which ideal MHD breaks down.
\item
Examining the scaling of the current layer thickness with braid complexity, one can extrapolate to place an upper bound on the braid complexity possible at coronal plasma parameters. For the studied magnetic field expression, Eq.~(\ref{beq}), this corresponds to $k\approx 3$. This threshold level of braid complexity for reconnection onset in the corona could be considered as a refinement to the notion of a critical angle between adjacent strands in a coronal loop \citep{dahlburg2005}.
\item
At this threshold braiding level, the excess magnetic energy available for release is in the range $10^{26}-10^{28}{\rm ergs}$, for $B_0$ in the range $10-100{\rm G}$. This is well above the energy range of nanoflares in typical models. However, a complete relaxation is not expected, and in reality the size of energy release events and  overall heating rate will then be determined by a balance between the reconnection-mediated energy release that acts to untangle the field, and the photospheric driving that typically acts to increase the braid complexity. Moreover, we have used here a braiding pattern that is close to optimal for efficiently tangling all field lines in the tube. For a less efficient braiding one might expect a turbulent non-ideal relaxation only in some sub-volume, releasing a smaller fraction of the total energy.
%This suggests that this simple model of braiding of a homogeneous field could explain the diffuse background corona, but may not provide sufficient energy to explain hot loops in the active corona. However, this may be improved by including additional complexity present on the Sun, such as the complex topological structure induced by the myriad photospheric flux sources, atmospheric stratification, and so on.
%\item
%The field perpendicular to the tube axis has peak value of order $25\%$ of the axial field strength even for the moderate levels of braiding addressed here, and at the predicted upper bound for reconnection onset in the corona, we estimate that $B_\perp$ will be of the same order as the axial field in the vicinity of the current layers. This calls into question the applicability of the reduced MHD approximation for studying braided coronal loops.
%\item
%The detailed pattern of the braiding is important for energy considerations. Different patterns of braiding will inject different quantities of energy into the field prior to the instability threshold being reached. The time to reach this threshold is dependent on the scaling of the QSL width with $k$, which in turn depends on the braiding pattern (see Section \ref{diffbraid}). Also, as pointed out by \cite{wilmotsmith2011b}, different fractions of the stored energy will be available for release in a turbulent relaxation for different braiding patterns.
\end{itemize}

Our results support the hypothesis that the corona can be heated through the braiding of the coronal field by turbulent boundary motions, that naturally lead to the formation of thin current layers in the corona. They also support the widely held view that these current layers do not form as spontaneous tangential discontinuities of $\BB$ (in the perfectly conducting limit), but rather that they become exponentially thin and intense as the stressing of the field proceeds. 
The results lead to a conclusion similar in spirit to the notion of the corona reaching a self-organised state governed by a critical angle between adjacent magnetic strands for reconnection onset \citep{dahlburg2005}. We propose that this critical angle should be replaced by a critical degree of field line braiding -- as measured by the thickness of layers in the squashing factor $Q$ in the domain.

Based on our results, studies such as that by \cite{yeates2012} can be used to estimate the timescale for the coronal field to becomes sufficiently braided for heating to occur. Specifically, they used local correlation tracking to measure flow velocities at the photosphere, and constructed the corresponding coronal field assuming a uniform field at the start of the observations and an ideal evolution in the volume. They showed that the squashing factor $Q$ increased in net value and developed progressively smaller scales with time. It will be important in future to perform similar studies in different regions of the corona with as high as possible spatial and temporal resolution to capture the complexity of the small-scale braiding motions, in order to estimate braiding timescales.
What role field line braiding plays in explaining the observed coronal temperatures in active and quiet regions eventually depends on these timescales for the braiding motions, and the dynamic balance that is set up with the energy release and plasma response. It should be noted that the model considered contains many simplifications that need to be addressed in future, not least of which is the true complexity of the coronal field, which penetrates the photosphere in discrete flux sources and is permeated by a complex web of topological structures such as null points, separatrix surfaces, and separators.

\acknowledgments
Financial support from the UK's STFC (grant number ST/K000993) and fruitful discussions with A.~Wilmot-Smith, S.~Candelaresi and P.~Wyper are gratefully acknowledged. DP also acknowledges financial support from the Leverhulme Trust. Computations were carried out on the UKMHD consortium cluster funded by STFC and SRIF.

\appendix

\section{Proximity to an ideal evolution}\label{appA}
\begin{figure}
\centering
(a)\includegraphics[width=5cm]{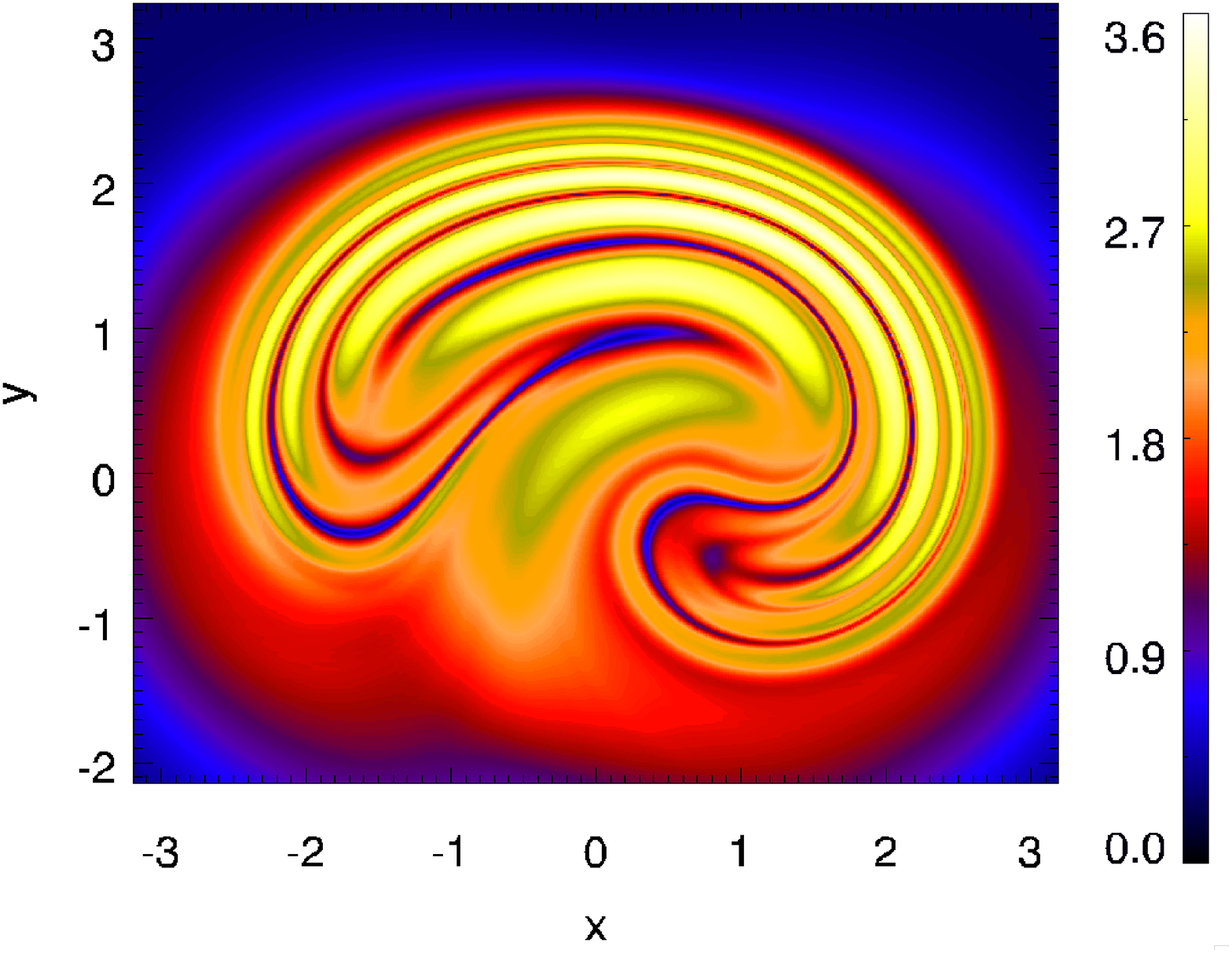}
(b)\includegraphics[width=5cm]{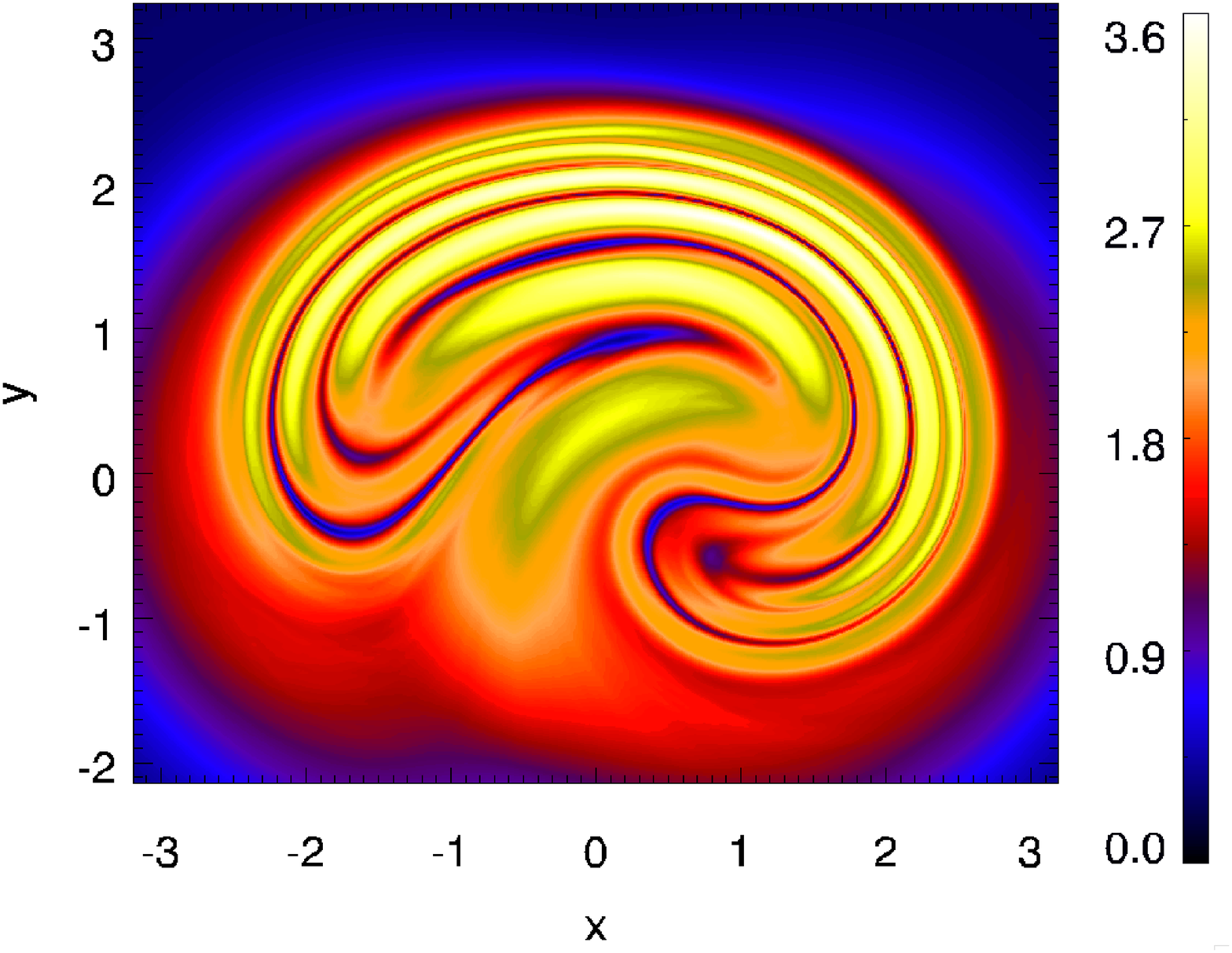}
\caption{Squashing factor $Q$ plotted on the lower boundary $z=-24$ for the simulation run with $k=0.6$, at (a) $t=0$, and (b) $t=300$.}
\label{qmapcomp}
\end{figure}

As discussed in section \ref{numsec}, the code used for the second stage of the relaxation -- the focus of this paper -- does not identically preserve the topology of the magnetic field. Although we set $\eta$ explicitly to zero, there will still be some non-zero numerical diffusion. However, as shown in Fig.~\ref{qmapcomp} the field line mapping, as measured by $Q$, is well preserved during the relaxation: there is very little change visible in the $Q$ profile that represents the field line mapping. To provide a more quantitative measure of the `idealness' of the relaxation, we can evaluate the net relative difference between the field line mapping at the initial time and in the relaxed state. Specifically, we calculate
\begin{equation}
\mathcal{E}=\frac{\displaystyle\int_D |\XX_0(x,y)-\XX_t(x,y)|}{\displaystyle\int_D |\XX_0(x,y)|}
\end{equation}
where $D$ is the horizontal domain shown in Fig.~\ref{qmapcomp}, and $\XX_0(x,y)=(X_0(x,y),Y_0(x,y))$ and $\XX_t(x,y)=(X_t(x,y),Y_t(x,y))$ are the field line mappings at $t=0$ and $t>0$, respectively. We evaluate $\mathcal{E}$ at $t=300$ for each simulation run ($480^2\times240$ resolution) -- by this time the field is in the asymptotic final state (see Figs.~\ref{j_lor_ev},\ref{minwidth_p6p7}).
For $k=0.5,0.55,0.6,0.65,0.7,0.75$ we find, respectively, $\mathcal{E}=0.0067$, $0.0097$, $0.012$, $0.016$, $0.019$, $0.022$. That is, the relative error in the field line footpoint location is between $0.7\%$ and $2.2\%$ (at $t=200$ this relative error is between $0.5\%$ and $1.4\%$). This naturally increases with increasing $k$ since the current becomes larger, and the field line mapping becomes more sensitive. These figures demonstrate that the braid structure is well preserved during the relaxation.
%%%%%%%%%%%%
%%%%%%%%%%%%

\bibliographystyle{apalike} 
%\bibliography{bibliog}

\end{document}